\documentclass[twoside,twocolumn,9pt]{article}
\usepackage{extsizes}
\usepackage[super,sort&compress,comma]{natbib} 
\usepackage[version=3]{mhchem}
\usepackage[left=1.5cm, right=1.5cm, top=1.785cm, bottom=2.0cm]{geometry}
\usepackage{balance}
\usepackage{mathptmx}
\usepackage{sectsty}
\usepackage{graphicx} 
\usepackage{lastpage}
\usepackage[format=plain,justification=justified,singlelinecheck=false,font={stretch=1.125,small,sf},labelfont=bf,labelsep=space]{caption}
\usepackage{float}
\usepackage{fancyhdr}
\usepackage{fnpos}
\usepackage[english]{babel}
\addto{\captionsenglish}{%
  
}
\usepackage{array}
\usepackage{droidsans}
\usepackage{charter}
\usepackage[T1]{fontenc}
\usepackage[usenames,dvipsnames]{xcolor}
\usepackage{setspace}
\usepackage[compact]{titlesec}
\usepackage{hyperref}

\usepackage{amssymb}

\usepackage{subfig}
\usepackage{tabularx}
\usepackage{footnote}
\usepackage{threeparttable}

\usepackage{xcolor}
\usepackage{geometry}

\usepackage{epstopdf}%This line makes .eps figures into .pdf - please comment out if not required.

\definecolor{cream}{RGB}{222,217,201}

\begin{document}

\pagestyle{fancy}
\thispagestyle{plain}
\fancypagestyle{plain}{
%%%HEADER%%%
\renewcommand{\headrulewidth}{0pt}
}
%%%END OF HEADER%%%

%%%PAGE SETUP - Please do not change any commands within this section%%%
\makeFNbottom
\makeatletter
\renewcommand\LARGE{\@setfontsize\LARGE{15pt}{17}}
\renewcommand\Large{\@setfontsize\Large{12pt}{14}}
\renewcommand\large{\@setfontsize\large{10pt}{12}}
\renewcommand\footnotesize{\@setfontsize\footnotesize{7pt}{10}}
\renewcommand\scriptsize{\@setfontsize\scriptsize{7pt}{7}}
\makeatother

\renewcommand{\thefootnote}{\fnsymbol{footnote}}
\renewcommand\footnoterule{\vspace*{1pt}% 
\color{cream}\hrule width 3.5in height 0.4pt \color{black} \vspace*{5pt}} 
\setcounter{secnumdepth}{5}

\makeatletter 
\renewcommand\@biblabel[1]{#1}            
\renewcommand\@makefntext[1]% 
{\noindent\makebox[0pt][r]{\@thefnmark\,}#1}
\makeatother 
\renewcommand{\figurename}{\small{Fig.}~}
\sectionfont{\sffamily\Large}
\subsectionfont{\normalsize}
\subsubsectionfont{\bf}
\setstretch{1.125} %In particular, please do not alter this line.
\setlength{\skip\footins}{0.8cm}
\setlength{\footnotesep}{0.25cm}
\setlength{\jot}{10pt}
\titlespacing*{\section}{0pt}{4pt}{4pt}
\titlespacing*{\subsection}{0pt}{15pt}{1pt}
%%%END OF PAGE SETUP%%%

%%%FOOTER%%%
\fancyfoot{}
%\fancyfoot[LO,RE]{\vspace{-7.1pt}\includegraphics[height=9pt]{head_foot/LF}}
%\fancyfoot[CO]{\vspace{-7.1pt}\hspace{13.2cm}\includegraphics{head_foot/RF}}
%\fancyfoot[CE]{\vspace{-7.2pt}\hspace{-14.2cm}\includegraphics{head_foot/RF}}
\fancyfoot[RO]{\footnotesize{\sffamily{1--\pageref{LastPage} ~\textbar  \hspace{2pt}\thepage}}}
\fancyfoot[LE]{\footnotesize{\sffamily{\thepage~\textbar\hspace{3.45cm} 1--\pageref{LastPage}}}}
\fancyhead{}
\renewcommand{\headrulewidth}{0pt} 
\renewcommand{\footrulewidth}{0pt}
\setlength{\arrayrulewidth}{1pt}
\setlength{\columnsep}{6.5mm}
\setlength\bibsep{1pt}
%%%END OF FOOTER%%%

%%%FIGURE SETUP - please do not change any commands within this section%%%
\makeatletter 
\newlength{\figrulesep} 
\setlength{\figrulesep}{0.5\textfloatsep} 

\newcommand{\topfigrule}{\vspace*{-1pt}% 
\noindent{\color{cream}\rule[-\figrulesep]{\columnwidth}{1.5pt}} }

\newcommand{\botfigrule}{\vspace*{-2pt}% 
\noindent{\color{cream}\rule[\figrulesep]{\columnwidth}{1.5pt}} }

\newcommand{\dblfigrule}{\vspace*{-1pt}% 
\noindent{\color{cream}\rule[-\figrulesep]{\textwidth}{1.5pt}} }

\makeatother
%%%END OF FIGURE SETUP%%%

%%%TITLE AND AUTHORS%%%
\twocolumn[
  \begin{@twocolumnfalse}
\vspace{1em}
\sffamily

\LARGE{\textbf{Plasmonic and metamaterial biosensors: A game-changer for virus detection}}\\

\begin{tabular}{m{4.5cm} p{13.5cm} }

%Article title goes here instead of the text "This is the title"
 & \vspace{0.3cm} \\

 & \noindent\large{Junfei Wang,\textit{$^{a}$} Zhenyu Xu,\textit{$^{a}$} and Domna G. Kotsifaki$^{\ast}$\textit{$^{a,b}$}} \\%Author names go here instead of "Full name", etc.

\end{tabular}

 \end{@twocolumnfalse} \vspace{0.6cm}

  ]
%%%END OF TITLE AND AUTHORS%%%

%%%FONT SETUP - please do not change any commands within this section
\renewcommand*\rmdefault{bch}\normalfont\upshape
\rmfamily
\section*{}
\vspace{-1cm}

%%%FOOTNOTES%%%

\footnotetext{\textit{$^{a}$~Division of Natural and Applied Sciences, Duke Kunshan University,  Kunshan, 215316, Jiangsu Province, China. E-mail: domna.kotsifaki@dukekunshan.edu.cn}}
\footnotetext{\textit{$^{b}$~Data Science Research Center, Duke Kunshan University, Kunshan,215316, Jiangsu Province, China.}}

%%%END OF FOOTNOTES%%%

%%%ABSTRACT%%%%

\sffamily{\textbf{One of the most important processes in the fight against current and future pandemics is the rapid diagnosis and initiation of treatment of viruses in humans. In these times, the development of high-sensitivity tests and diagnostic kits is an important research area. Plasmonic platforms, which control light in subwavelength volumes, have opened up exciting prospects for biosensing applications. Their significant sensitivity and selectivity allow for the non-invasive and rapid detection of viruses. In particular, plasmonic-assisted virus detection platforms can be achieved by various approaches, including propagating surface and localized plasmon resonances, as well as surface-enhanced Raman spectroscopy. In this review, we discuss both the fundamental principles governing a plasmonic biosensor and prospects for achieving improved sensor performance. We highlight several nanostructure schemes to combat virus-related diseases. We also examine technological limitations and challenges of plasmonic-based biosensing, such as reducing the overall cost and handling of complex biological samples. Finally, we provide a future prospective for opportunities to improve plasmonic-based approaches to increase their impact on global health issues.}}\\

%%%END OF ABSTRACT%%%%

\rmfamily %Please do not remove this line.

%%%MAIN TEXT%%%%

\section{Introduction}
At the dawn of twenty-first century, humanity faces multiple health challenges with substantial global economic and social impacts~\cite{Soler,Giovannini, Mohammad, Asghari, Hatice, Serafetinides}. The monitoring and early detection of biological entities necessitates platforms that are able to analyze extremely low concentrations of analytes in real samples near the point of care (PoC) and sometimes at the place of patient care. The early detection and timely treatment of diseases can improve cure rates and reduce treatment costs. Commonly used analytical methods~\cite{Caliendo,Boonham,Kaiser} rely upon culture-based methods, serological tests, or nucleic acid-based amplification techniques such as polymerase chain reaction (PCR), gene sequencing, virus isolation, hemagglutination assay, and enzyme-linked immunosorbent assay (ELISA). In spite of their inherent advantages, these techniques are time consuming and involve sophisticated instrumentation that requires skilled operators. In addition, time-consuming predeveloped protocols are typically limited to specific strains or types of viruses and may have high false-negative rates, which limit their effectiveness to lower the risk of new infections~\cite{Cui}. Consequently, the need for new diagnostic approaches that are fast and cost effective has brought into focus the development of real-time PoC testing diagnostic devices~\cite{Menendez}, which could be game-changers for the management of diseases. 

With the growing need for PoC diagnostic platforms, the World Health Organization has created the ASSURED (affordable, sensitive, specific, user-friendly, rapid and robust, equipment free, and deliverable to end users) framework, outlining directions and guidelines for their development~\cite{Mabey}. Current PoC tests, such as paper-based devices~\cite{Yetisen}, succeed in providing rapid, cost-effective, and facile results but are held back by inadequate sensitivity, selectivity, and overall reliability, highlighting the challenges faced by PoC diagnostics~\cite{Zhang_2019}. Early diagnosis is essential for a wide range of conditions, including infectious diseases, auto-immune disorders, and inflammatory diseases, for which timing is important to maximize the efficacy of therapy. In addition, continuous monitoring of biomarkers or therapeutic drug levels at the bedside can provide valuable feedback to physicians and allow them to tailor the treatment options for individual patients~\cite{Mage,Visser,Jason}. In this aspect, nanostructure-based PoC approaches that can rapidly provide the molecular profile of a patient could become instrumental in paving the way towards precision diagnosis~\cite{Minhaz,HO}.

%%%%%%%%%%%%%%%%%%%%%%%%%%%%%%%%%%%%%%%%%%%%%%%%%%
%Figure 1
%%%%%%%%%%%%%%%%%%%%%%%%%%%%%%%%%%%%%%%%%%%%%%%%%%
\begin{figure*}[ht]
\centering
\includegraphics[trim={0cm 0cm 0cm 0cm},clip, width=1\textwidth]{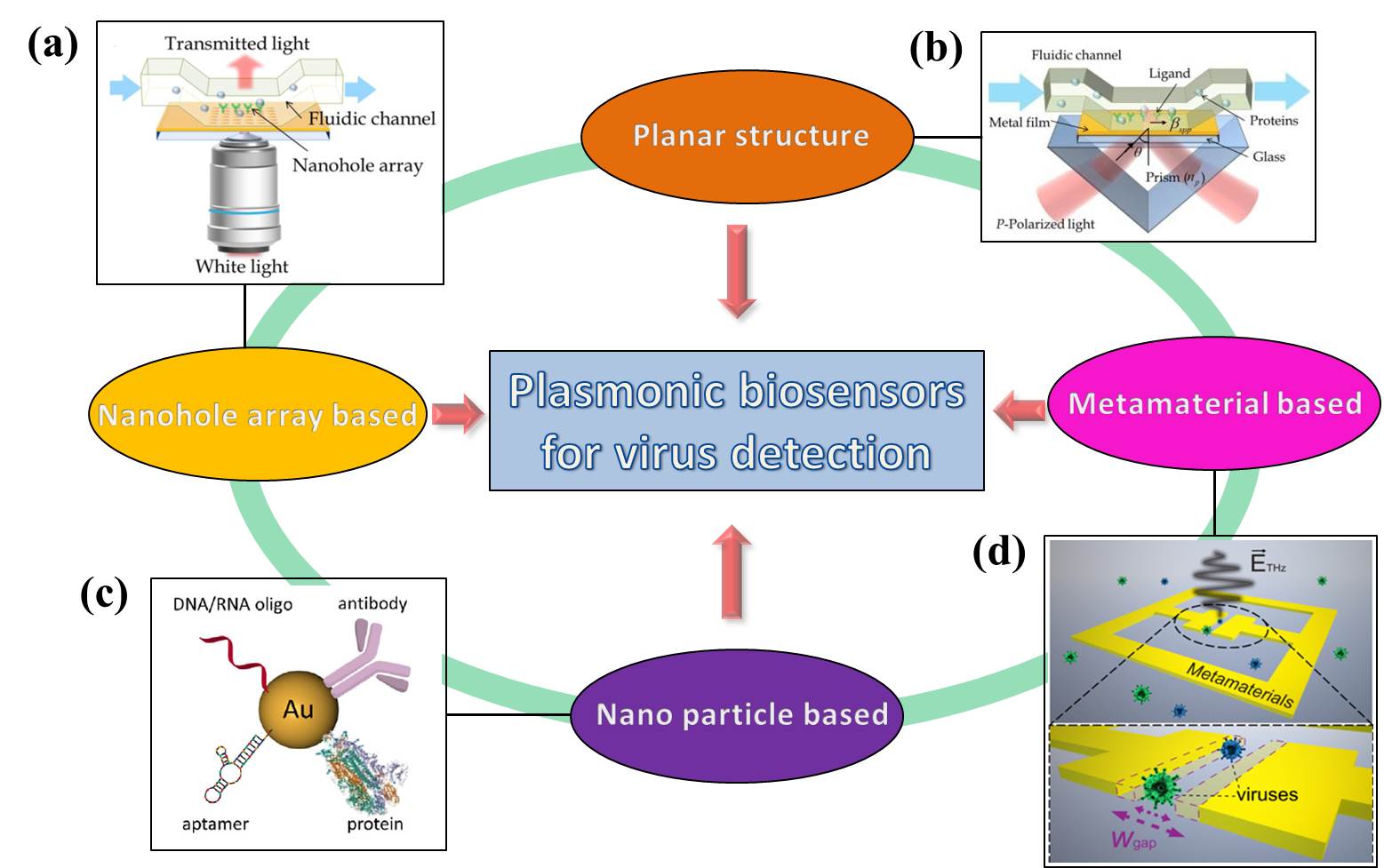}
\setlength\abovecaptionskip{10pt}
\caption{\label{Fig.1} Schematic illustration of various plasmonic based virus sensing platforms. (a) An array of nanoholes can increase the binding potential for flowing virus antigens and enhance the sensitivity through the extraordinary transmission effect~(reproduced with permission from~\cite{Roh}). (b) Planar structure in which surface plasmon is generated in between a dielectric and a metal~(reproduced with permission from~\cite{Roh}). (c) Localized surface plasmon around nanoparticles can increase sensitivity~(reproduced with permission from~\cite{Jianxin}). (d) Metamaterials can efficiently enhance the electromagnetic fields of light, leading to ultrasensitive biosensing~(reproduced with permission from~\cite{Park}).}  
\end{figure*}
%%%%%%%%%%%%%%%%%%%%%%%%%%%%%%%%%%%%%%%%%%%%%%%%%%%%%%%

Plasmonic-based biosensing (Fig.~\ref{Fig.1}) have embraced the challenge of offering on-site strategies to complement traditional diagnostic methods and has attracted significant attention owing to its versatility and abillity to acheive label-free monitoring with low response times~\cite{Minhaz,Zihan,Lee,Sile1,Hatice,QU,Das,Mauriz,sile2}. These characteristics, achieved by exploiting the properties of nanomaterials~\cite{Krejcova,Adegoke,Laspada,Pang,EPA,abdel}, has allowed for the design of ultrasensitive nanobiosensors, which could be implemented in diagnostic tools to alleviate the burden of infectious diseases in the developing world. Moreover, as light sources, detectors, and optical components are abundant in the visible-to-near infrared electromagnetic spectrum range, the design of plasmonic biosensors in this range is particularly advantageous~\cite{Roh}. Such biosensors require structural dimensions on the few-nanometer scale and can be fabricated using today’s nanolithography techniques~\cite{Zhang_2019,Roh,Abbas,Yesudasu}. Furthermore, plasmonic biosensors enable direct detection of analytes from heterogeneous biological media without the need of exogenous labels~\cite{Li,Couture}. This is a key factor in plasmonic-based biosensor design since it facilitates bio-assay procedures by eliminating tedious washing, amplifying, and labeling steps~\cite{Soler,Hasan,Cui}. For these reasons, plasmonic-based biosensors are seen as promising candidates for the essential elements of future biosensor PoC platforms.

In this tutorial review, we present the advances in plasmonic-based biosensing for virus detection and highlight the scope of future work in this research field. We address the fundamental physical principles of plasmonic effects and biosensing strategies. The integration of metallic nanostructures into commercial microfluidic platforms for future devices that can alert the public to biological threats is also discussed. Because of the ongoing coronavirus disease 2019 (COVID-19) pandemic, slight emphasis is given to coronavirus detection techniques. Finally, we discuss the challenges that need to be overcome for the future development of plasmonic-based biosensors and note how such biosensors are already impacting the diagnosis of infectious diseases in the developing world. We believe that this comprehensive review will be a useful resource for researchers, physicians, and students interested in constructing ultra-dense and high-throughput clinical screening plasmonic devices.
 
%%%%%%%%%%%%%%%%%%%%%%%%%%%%%%%%%%%%%%%%%%
\section{Physical Considerations}

\subsection{A brief historical introduction}

The interaction of light with plasmonic nanostructures has long been a subject of interest in the classical and quantum worlds~\cite{novotny_hecht_2012}. A key feature of plasmon resonances is that they are excited by electromagnetic waves, either evanescent or localized~\cite{novotny_hecht_2012}. Their first observation dates back to Wood~\cite{Wood}, who reported anomalous reflective patterns when polarized light was shone on a metallized diffraction grating. A few years later, Rayleigh~\cite{Rayleigh} provided a phenomenological explanation for these patterns, but the underlying physical mechanism remained a mystery. In 1957, significant advances in our understanding of surface plasmon resonance (SPR) were made when Ritchie ~\cite{Ritchie} confirmed the presence of metal surface plasma excitations, while Powell~\textit{et al.}~determined that the excitation of surface plasmons involved electrons at metal interfaces~\cite{Powell}. In 1968, Otto used an attenuated total reflection prism-coupled method to enable the coupling of an electromagnetic field with surface plasmon waves~\cite{Otto}. Similarly, Kretschmann and Raether reported the excitation of SPR by utilizing a 10~–~100~nm thin gold film on the surface of a prism~\cite{Kretschmann}.
%%%%%%%%%%%%%%%%%%%%%%%%%%%%%%%%%%%%%%%%%%%%%%%%%%
%Figure 2
%%%%%%%%%%%%%%%%%%%%%%%%%%%%%%%%%%%%%%%%%%%%%%%%%%
\begin{figure*}[h]
\centering
\includegraphics[trim={0.5cm 0.5cm 1.5cm 0.3cm},clip, width=0.9\textwidth]{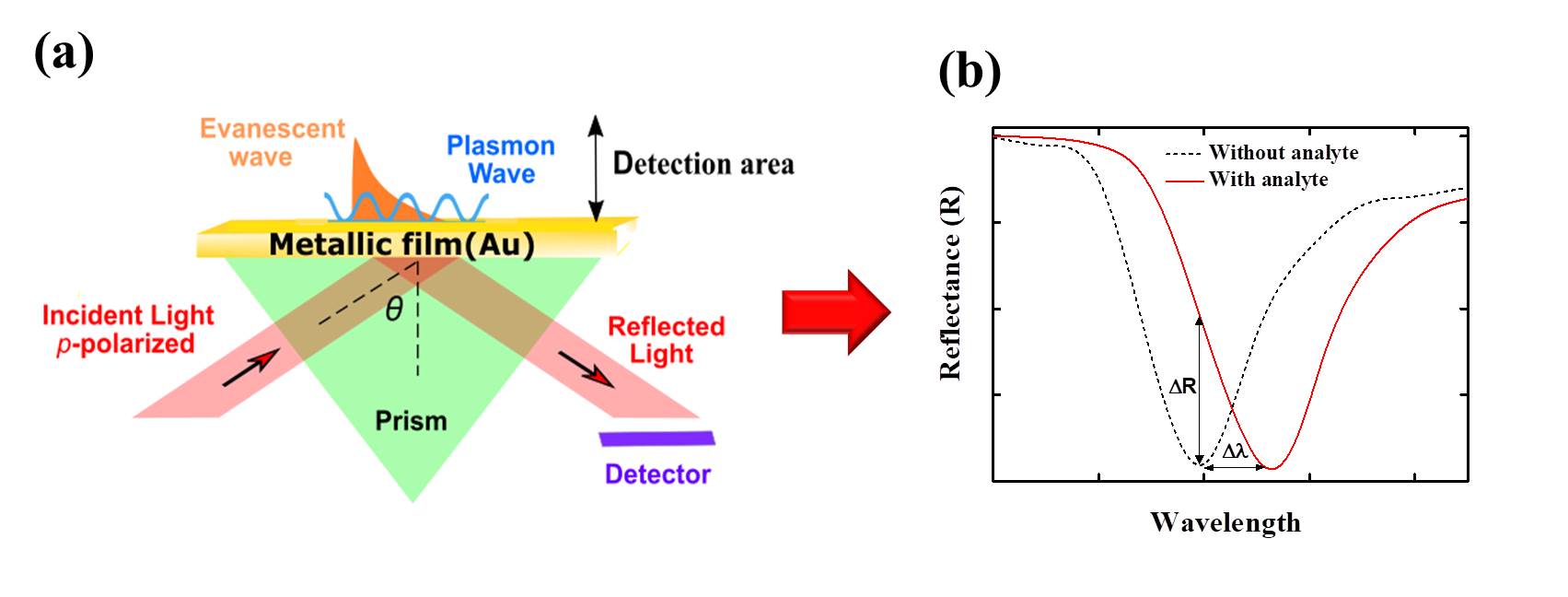}
\setlength\abovecaptionskip{0pt}
\caption{\label{Fig.2} (a) Schematic illustration of the Kretschmann configuration~\cite{Bouloumis}. The surface plasmon polariton can be excited when momentum mismatching is satisfied. The surface plasmon decays exponentially from the surface and propagates to a distance of a few tens of microns. (b) Typical sensor readouts: spectrum of reflected light before and after the binding of analyte, which leads to refractive index changes. The momentum mismatching condition exists at certain incident angles~$\theta$. $\Delta$$\lambda$ is the resonance shift and $\Delta$R indicates the intensity changes due to analyte binding.}  
\end{figure*}
%%%%%%%%%%%%%%%%%%%%%%%%%%%%%%%%%%%%%%%%%%%%%%%%%%%%%%%
The potential exploitation of SPR for biosensing first appeared in 1970 with observations made by Fleischmann and colleagues~\cite{Fleischmann}, who noted an enhancement of Raman scattering near a roughened metal surface; this enhancement was later found to be associated with an electromagnetic effect~\cite{jeanmaire}. Ten years later, Liedberg~\textit{et al.} observed refractive index (RI) changes on the surface of a metallic film after the absorption of biomolecules~\cite{Liedberg}. Since then, the label-free nature of SPR biosensing has become an important tool in biophysics, molecular biology, and pharmaceutical research~\cite{Simona, Abbas, Saeideh, Bockov,Yesudasu}. Today, several companies, such as Biocore, PhotonicSys, and Plasmetric, manufacture devices used to evaluate the performance of biosensor chips for PoC applications.  

\subsection{Fundamental mechanism of plasmon resonance biosensors}

\subsubsection{SPR mechanism}

Surface plasmon polaritons require a material with free electrons and low optical losses at the optical regime of the electromagnetic wave,~\textit{i.e.}, materials that possess a negative real and small positive imaginary dielectric constant~\cite{Willets,Johnson,Yang}. Among the materials that exhibit plasmonic properties, conductive noble metals, such as gold and silver, are often used to excite plasmon resonances because of their tunable plasmonic properties in the visible and near-infrared range of the electromagnetic spectrum~\cite{Halas}. 

A simple technique of generating surface plasmons from a metal-dielectric interface is the Kretschmann configuration (Fig.~\ref{Fig.2}(a)). The underlying physics of SPR sensors based on an evanescent field has been reviewed extensively in literature~\cite{Raether,Ming,Couture}. Briefly, when light, which implies an electromagnetic wave, strikes the metal, the electric field of the light interacts with conducting electrons. The coupling of the incident electromagnetic wave to the collective oscillations of the conduction electrons forms an evanescent wave, which is known as  SPR. To achieve this, the momentum of incident photons should match the momentum of the conduction band of electrons. This momentum matching condition depends on the RI of the dielectric medium at the surface of the metal layer and is given by the following expression~\cite{Raether}:

\begin{equation}
    k_{SP} = \frac{2\pi}{\lambda n} \sqrt{\frac{\varepsilon_{m}\varepsilon_{d}}{\varepsilon_{m}+\varepsilon_{d}}} = \frac{2\pi}{\lambda} n_{p} sin \theta
    \label{Eq.1}
\end{equation}

\noindent where~\textit{$k_{SP}$} is the surface plasmon wavevector, $\lambda$ is the wavelength of incident light, $\varepsilon_{m}$ is the dielectric constant of the metalic film (a function of $\lambda$), $\varepsilon_{d}$ (a function of the refractive index of the medium) is the dielectric constant of the surrounding medium,~\textit{$n_{p}$} is the refractive index of the coupling prism, and~\textit{$\theta$} is the incident angle of the light. When the $\varepsilon_{m}$ and $\varepsilon_{d}$ have equal magnitude and opposite sign the wavevector,~\textit{$k_{SP}$}, is maximum which results in plasmon resonance conditions. At this matching condition, the reflected light has minimum intensity and \textit{$\theta$} is called SPR angle. In particular, the evanescent field is highly sensitive to the refractive index (RI) of the analyte medium cause plasmonic properties changes (Fig.~\ref{Fig.2}(b)). Therefore, measuring the changes in this resonance condition (for example, angle, wavelength, intensity or phase), the biomolecular interactions that occur at the biosensor surface can be monitoring in real-time~\cite{Alessandra}.  

\subsubsection{Localized surface plasmon resonance (LSPR) mechanism}

LSPR is another mechanism that has potential applications in high-sensitivity plasmonic biosensing~\cite{Kravets,Halas,EPA}. Unlike in SPR, the electromagnetic field in LSPR does not propagate but is instead localized around subwavelength nanoparticles or nanostructures bound to the metal~\cite{Roh,Kravets}. The practical application of LSPR can be seen in artifacts dating back to the fourth century, with the Lycurgus Cup, an ancient Roman cage cup (currently on display in the British Museum~\cite{Ian}), being one of the best early examples of LSPR. The vessel is made of glass containing silver nanoparticles, leading to a green appearance when viewed with reflected light but a red appearance when viewed with transmitted light (from inside the cup). Specifically, the conduction electrons in the metallic nanoparticles undergo collective harmonic oscillations when under an applied electromagnetic field, resulting in a dipolar response~\cite{Li}. For a metallic nanoparticle with radius~\textit{R} and dielectric constant $\varepsilon_{m}$, in a medium with dielectric constant of $\varepsilon_{d}$, the exact conditions for LSPR can be solved by applying the Mie theory~\cite{Li}: 

\begin{equation}
    \sigma_{ext} = 12(\frac{\omega}{c}) \pi \varepsilon_{d}^{3/2} R^{3} \frac{Im(\varepsilon_{m})}{[Re(\varepsilon_{m})+2\varepsilon_{d}]^{2}}+[Im(\varepsilon_{m}]^{2} 
    \label{Eq.2}
\end{equation}

\noindent where $\omega$ is the angular frequency,~\textit{Im} and~\textit{Re} is the imaginary and real parts of dielectric constants, respectively. Equation~\ref{Eq.2} shows that when the electrons in the metallic nanoparticle oscillate and the real part of the dielectric function is negative, the denominator will vanish, which leads to a strong resonance condition that will shift with local changes in the dielectric environment~\cite{Ming}. In addition, coherent oscillations of the electrons at resonance make the absorption and scattering cross-sections several orders of magnitude larger than the actual size of the nanoparticles~\cite{Ming}. Equation~\ref{Eq.2} is modified with the geometrical form factor, while for any arbitrary shape, more rigorous calculations are needed~\cite{EPA}. Moreover, the performance of LSPR biosensors depends on the resonance properties of the nanostructures, which can be engineered by optimizing the design parameters. For instance, nanorods with a high aspect ratio are more sensitive to RI changes~\cite{Halas} while larger metallic nanoparticles have smaller repulsion of electrons at opposite surfaces, resulting in plasmon resonance that is more red-shifted; this is suitable for making LSPR-biosensors that can detect and quantify biorecognition events~\cite{Chen}. Additionally, the incident electromagnetic light can be directly coupled to the plasmon field without any coupling configuration, such as with prisms or gratings, which improves the complexity of the sensing system. Furthermore, LSPR-based biosensor nanostructures can be fabricated by nanolithography techniques using not only nanoparticles but also chip-based substrates that are miniaturized with high sensitivity and repeatability. This can provide the benefit of being able to integrate the biosensor with other sensing components, such as microfluidics~\cite{Brolo, Srdjan, Mackenzie}.  

For the detection mechanisms in both SPR and LSPR, the sensitivity to changes within their associated plasmon decays with length~\cite{Ming}. LSPR changes can be detected within tens of nanometers in the visible range, whereas SPR changes, which occur along the propagation surface, can be detected within a few hundred nanometers~\cite{Kravets}. In biosensing, LSPRs are usually utilized through surface-enhanced techniques such as surface-enhanced Raman scattering spectroscopy ~\cite{Moskovits} (SERS), surface-enhanced infrared absorption spectroscopy~\cite{Kundu}, surface-enhanced fluorescence~\cite{Fayyaz}, and through resonance shifts induced by nearby analytes~\cite{Taylor}.

\subsection {Plasmonic metamaterials}

Plasmonic metamaterials have been utilized to further control collective plasmonic modes and electromagnetic field enhancement~\cite{Boris, Arash, Kotsifaki_Fano, Kotsifaki_Fano2}. The concept of these materials was first introduced in 1968 by Veselago, who observed the unusual behavior of light refracted by a left-handed material~\cite{Veselago}. A few years later, Pendry~\textit{et al.} noted that microstructures, fabricated from nonmagnetic conducting sheets, smaller than the excitation wavelength could be tuned to show varying magnetic permeability, including imaginary components~\cite{Pendry}. Based on these observations~\cite{Veselago,Pendry}, a practical way to manufacture a left-handed material that does not follow the conventional right-hand was determined. In 2000, Smith~\textit{et al.} demonstrated the first left-handed metamaterial, which exhibited negative permeability and permittivity simultaneously at microwave frequencies~\cite{Smith}. Since then, metamaterials have been explored extensively for a variety of applications in optics~\cite{Pendry1}, photonics~\cite{Fang}, energy harvesting~\cite{Yu}, sensing~\cite{Beruete},imaging~\cite{Roh1}, and spectroscopy~\cite{Zhou}. 
Compared with conventional SPR-based methods, metamaterials can be more easily fabricated through nanolithography techniques~\cite{Arash}. For periodic arrays of metamolecules, near- and far-field coupling is utilized to generate resonance with a high quality factor (Q-factor). This breaks the damping limit of a single metamolecule in the dipole approximation~\cite{Odom}, thus making such arrays promising candidates for biosensing applications~\cite{Arash}. 

\subsection{SERS mechanism}

SERS is a highly analytical tool~\cite{FAN20201,D1NA00237F} that has many applications in the field of diagnostics~\cite{Bantz}. It can be used to enhance weak Raman signals of analytes through the use of plasmonic nanostructures~\cite {Le, Anderson, Moore}. Raman spectroscopy evaluates the vibrational and rotation modes of biomolecules through the analysis of inelastic Raman scattering of a laser beam~\cite{Le}. Specifically, metallic nanostructures possess a localized electromagnetic field as a result of LSPR, which affects the Raman signal of an active analyte in close proximity to the nanostructure by enhancing the Raman scattering cross-section~\cite{Anderson}. Overall, SERS shows a broad range of benefits, such as high selectivity due to the unique fingerprint signatures of analytes, easy sample preparation, high possibility of single-entity detection, high throughput, and PoC applicability by using available Raman probes~\cite{FAN20201,D1NA00237F}.

\subsection{General characteristics of plasmonic biosensors}

\begin{figure*}[h!]
\centering
\includegraphics[trim={0cm 0cm 0cm 0cm},clip, width=0.7\textwidth]{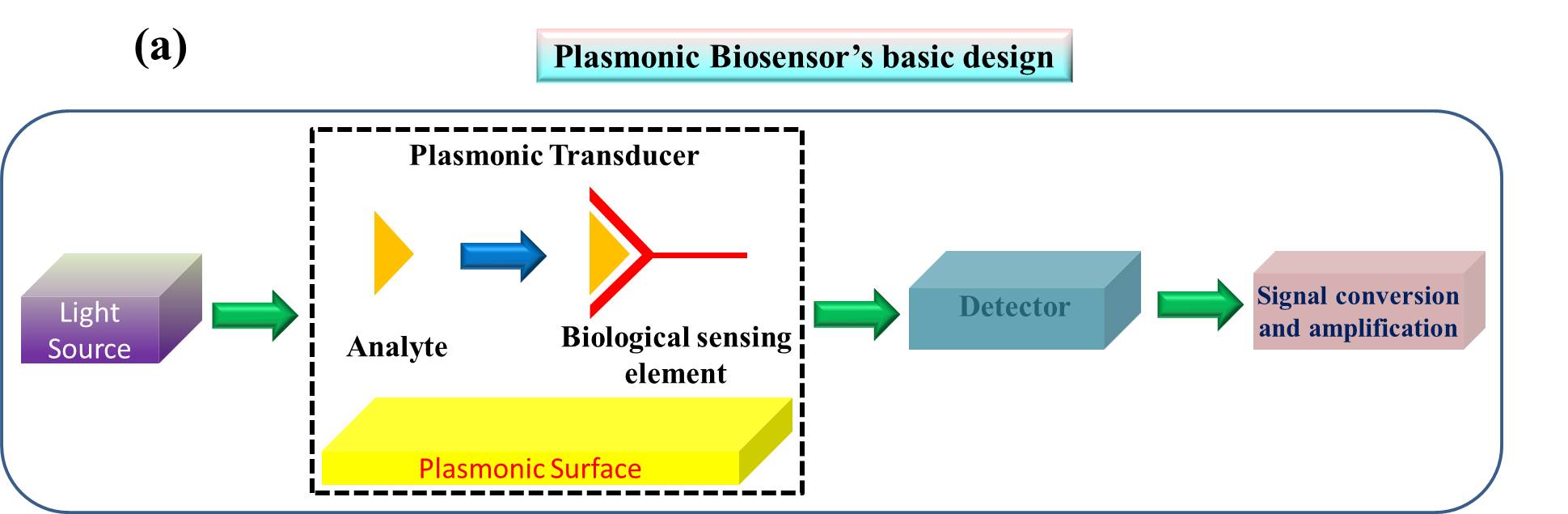}

\vspace{1cm}

\includegraphics[trim={0cm 0cm 0cm 0cm},clip, width=0.7\textwidth]{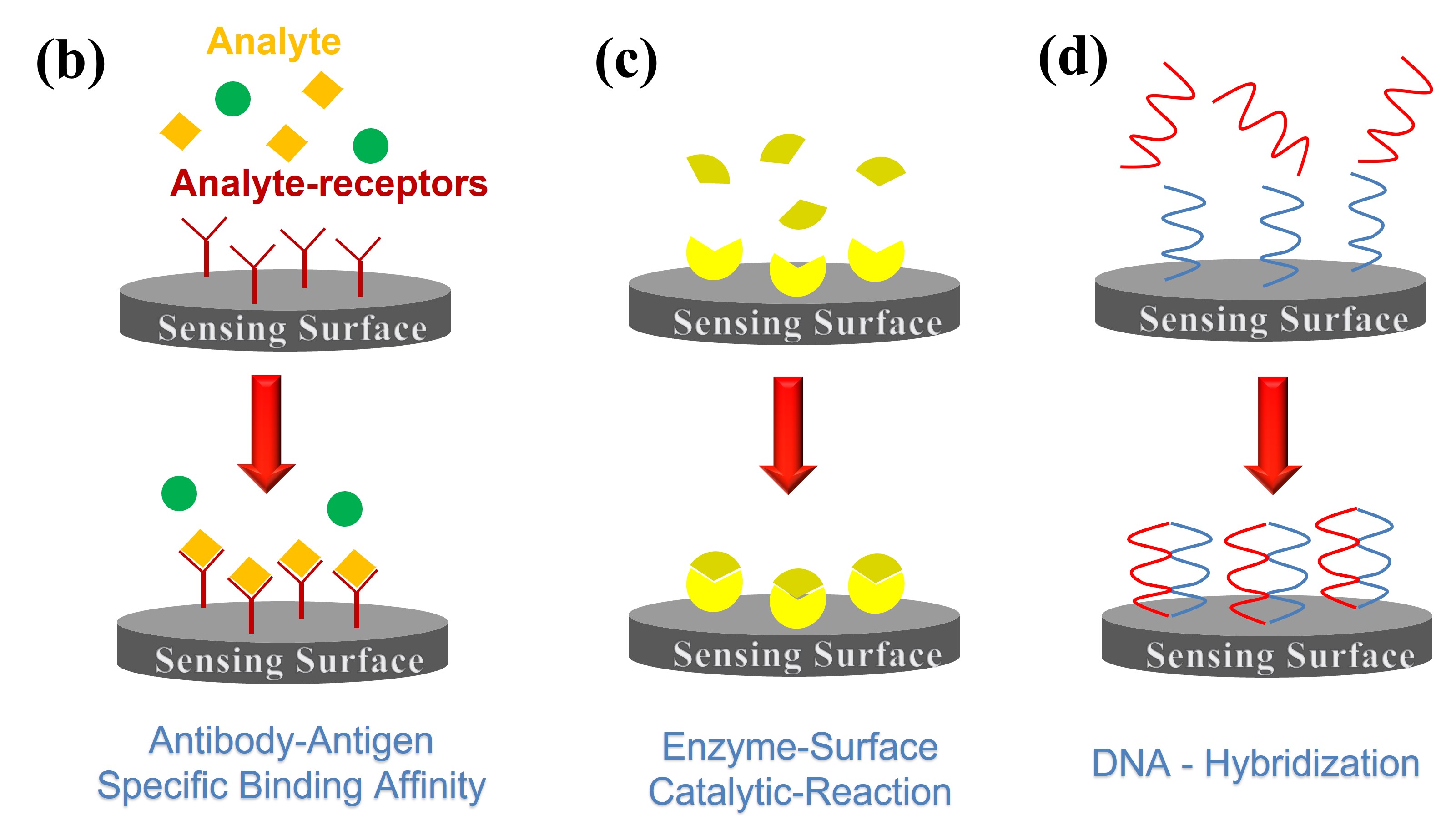}
\setlength\abovecaptionskip{0pt}
\caption{\label{Fig.3} (a) Schematic illustration of a plasmonic biosensor that translates the capture of the analyte to a measurable alteration of light intensity or resonance shift. Analyte-receptor coupling mechanisms on plasmonic biosensor surfaces include (b) antibody-antigen binding, (c) enzyme-surface catalytic reactions, and (d) DNA hybridization.}  \end{figure*}

The basic components of a biosensor are illustrated in Figure~\ref{Fig.3}(a) and consist of the target analyte bound to the bioreceptor, the transducer, which converts the signal into a measurable quantity, and the reader device, which generates the results ~\cite{Estrela,Huang1} (detailed descriptions of these components are available elsewhere and are beyond the scope of this work~\cite{Huang1,Estrela}). In addition, chemical activation of the surface is crucial to improve the sensing efficiency to single virus particles. Some important features of analyte-receptor coupling on the plasmonic surface are shown in  Figure~\ref{Fig.3}(b). Typically, in affinity-based plasmonic biosensors, surfaces are activated by biological receptors, such as antibodies, nucleic acids, cell membrane receptors, specifically designed peptides, aptamers, or molecularly imprinted polymers (MIPs)~\cite{Kozitsina}. 
%%%%%%%%%%%%%%%%%%%%%%%%%%%%%%%%%%%%%%%%%%%%%%%%%%
%Figure 3
%%%%%%%%%%%%%%%%%%%%%%%%%%%%%%%%%%%%%%%%%%%%%%%%%%
%%%%%%%%%%%%%%%%%%%%%%%%%%%%%%%%%%%%%%%%%%%%%%%%%%%%%%% 
These biological entities show great affinity and specificity for certain analytes, allowing for the selective capture of the target with high sensitivity from complex biological samples. In the following section, we will focus on several parameters used to assess the performance of a biosensor~\cite{Estrela}. 
In the context of biosensing, the most important feature of a sensor is its sensitivity,~\textit{S},~\cite{Estrela}; it is described by Equation 3 and can be determined by the slope of the analytical calibration curve,

\begin{equation}
    S = \frac{\Delta \lambda}{\Delta n}
    \label{Eq.3}
\end{equation}

\noindent where $\lambda$ is the surface plasmon resonance and~\textit{n} is the refractive index of the medium in contact with the sensor surface. The magnitude of a sensor’s sensitivity depends on the supporting electromagnetic mode, resonant wavelength, excitation geometry, and properties of the substrate~\cite{Barbora}. Therefore, bulk and surface sensitivities are not necessarily linked to each other. For instance, for a thin gold film excited at a low angle using a Kretschmann configuration, a high bulk sensitivity (>5,000~nm/RI unit [RIU]) can be achieved, while a simultaneous small surface sensitivity is obtained because of the high decay depth~\cite{Couture}. This implies that a small amount of an analyte (<10 kDa) can be easily detected using a biosensor with a small penetration depth~\cite{Couture, Miller}. 

Another key parameter is the limit of detection (LOD) or sensor resolution, which is defined as the smallest amount of analyte that can be reliably detected by a specific measurement process. It is determined by the concentration of the analyte that produces a biosensor response corresponding to the standard deviation,~\textit{$\sigma_{blank}$}, of the biosensor response measured with no analyte and is given by~\cite{Kolomenskii,Barbora}:

\begin{equation}
    LOD = m \frac{\sigma_{blank}}{S}
    \label{Eq.6}
\end{equation}

\noindent where~\textit{m} is a numerical factor. Typical resolution of $10^{-6}$~RUI have been demonstrated with gold films and a Kretschmann configuration~\cite{Live}. Piliarik and Homola~\cite{Piliarik} calculated the ultimate theoretical resolution of a SPR sensor to be $10^{-7}$~RUI. The authors showed that such a resolution could be reached regardless of the type of SPR coupling or signal modulation by, for example, increasing the signal-to-noise ratio of the detected light using high-end optoelectronic components~\cite{Piliarik}.

The performance of a plasmonic biosensor is strongly influenced by the spectral shape and background noise of the readout system. For the spectral shape, the \textit{Q}-factor is an important parameter since it is a reliable indicator of the resolution of the detector for certain analytes and is given by~\cite{Bergman}:

\begin{equation}
    Q = \frac{\lambda}{FWHM}
    \label{Eq.5}
\end{equation}
\noindent To enhance sensing performance, large \textit{Q}-factor values are desirable because sharper resonance peaks with large Q-factors make it much easier to detect small RI changes~\cite{Boris}. For example, nanostructures that support Fano resonances lead to sharp asymmetric peaks that show up to two-fold sensitivity enhancement when compared with conventional biosensors~\cite{Ahmet,Zhang}. 

Another important characteristic of a biosensor is the figure of merit (FOM), which is the ratio between the sensitivity and full width at half maximum of the resonance spectra~\cite{Miller}.  

\begin{equation}
    FOM = \frac{S}{FWHM}
    \label{Eq.4}
\end{equation}

\noindent The FOM is a key factor for evaluating and comparing different plasmonic nanostructures with respect to their sensing potential and is dependent on the metal film, prism material, and resonance~\cite{Offermans}. In conclusion, the optimum performance of a plasmonic biosensor should be evaluated after taking into account several factors that require carefully consideration.  

%%%%%%%%%%%%%%%%%%%%%%%%%%%%%%%%%%%%%%%%%%%%%%%%%%%%%%%%%%%%%%
\begin{table*}[h]
\centering
\caption{\textbf{Overview of plasmonic~-~based biosensors for virus detection}}
\label{tab:Table1}
\resizebox{\textwidth}{!}{%
\begin{threeparttable}[b]
\resizebox{\textwidth}{!}{\begin{tabular}{cccccc}
\hline \hline
\textbf{Structure} & \multicolumn{1}{l}{\textbf{Virus Detected}} &  
\textbf{\begin{tabular}[c]{@{}c@{}}Detection Format\\ \end{tabular}} &
\textbf{\begin{tabular}[c]{@{}c@{}} LOD\\ Sensitivity \end{tabular}} &
\textbf{\begin{tabular}[c]{@{}c@{}}Reference\\  \end{tabular}} \\
\hline \hline
\\
\begin{tabular}[c]{@{}c@{}} Ag/Au (35~nm/10~nm) chips \\ [0pt] 
   \end{tabular} & Avian influenza H7N9 & Monoclonal antibody (IgM) & 144~copies/mL&~\cite{Chang} \\ 
\\   
\begin{tabular}[c]{@{}c@{}} Cr/Ag/Au (3~nm/40~nm/10~nm) chips \\ [6pt] \end{tabular} & Human enterovirus 71 & Enterovirus antibody & 67~virus particles (vp)/mL &~\cite{Prabowo1}\\ 
\\
\begin{tabular}[c]{@{}c@{}} Cr/Au (2.5~nm/47~nm) chips \\ [6pt] \end{tabular} & H1N1, RSV, Adenovirus, SARS & PCR amplified viral bodies   & 0.5~nM for adenovirus/2~nM for SARS  &~\cite{Shi}\\
\\
\begin{tabular}[c]{@{}c@{}} Au SPR chip \\ [6pt] \end{tabular} & Ebola virus  & Monoclonal antibodies & 0.5~pg/mL &~\cite{Sharma} \\
\\
\begin{tabular}[c]{@{}c@{}} Biacore X bare gold chip \\ [6pt] \end{tabular}    & HIV & Hairpin DNA, capture probes & 48~fM &~\cite{Diao}\\
\\
\begin{tabular}[c]{@{}c@{}} Spreeta 2000 (S2k) chips with Au surfaces \\ [6pt] \end{tabular}  & Tuberculosis (TB) virus & antibody to Ag85-TB secretory protein & 10~ng/mL &~\cite{Trzaskowski}\\
\\
\begin{tabular}[c]{@{}c@{}} Array of Au nanoprismn\\ [6pt] \end{tabular} & Rotavirus &  Rotavirus capsid (2B4) antibody & 126~$\pm$ 3 PFU/mL &~\cite{Rippa}\\
\\
\begin{tabular}[c]{@{}c@{}} Array of Au nanodiscks and nanodots  \\ [6pt] \end{tabular}  & Ebola virus & A/C protein  & 220~fg/mL &~\cite{Zhang_2019} \\
\\
\begin{tabular}[c]{@{}c@{}} Planar toroidal gold metamaterial \\ [6pt] \end{tabular}& Zika virus  & Immobilized antibody & 5.81~GHz/log(pg/mL)   &~\cite{Ahmadivand1} \\
\\
\begin{tabular}[c]{@{}c@{}} Au toroidal metasensor\\ [6pt] \end{tabular} & SARS CoV-2 &  SARS antibody & 4.2~fmol &~\cite{Arash1}\\
\\
\begin{tabular}[c]{@{}c@{}} Au nanospikes\\ [6pt] \end{tabular} & SARS CoV-2 &  SARS antibody & 0.08~ng/mL &~\cite{Funari}\\
\\
\begin{tabular}[c]{@{}c@{}} Au nano-island layer\\ [6pt] \end{tabular} & SARS CoV-2 &  Thiol cDNA receptor & 0.22~pM &~\cite{Qiu}\\
\\
\begin{tabular}[c]{@{}c@{}} Hetero-assembled Au nanoparticles layer\\ [6pt] \end{tabular} & Hepatitis B virus &  Hepatitis antibody & 100~fg/mL &~\cite{Jinwoon}\\
\\
\begin{tabular}[c]{@{}c@{}} Au spike-like nanoparticles \\ [6pt] \end{tabular} & Avian influenza virus &  DNA - Hemagglutinin binding aptamer & 1~pg/mL &~\cite{lee2019}\\
\\
\begin{tabular}[c]{@{}c@{}} Au($\sim$ 20~nm) particles\\ [6pt] \end{tabular} & Norovirus &  Norovirus recognizing affinity peptide & 9.9~copies/mL &~\cite{Heo}\\
\\
\begin{tabular}[c]{@{}c@{}} Bioconjugated Au nanoparticle (10~nm-15~nm)\\ [6pt] \end{tabular} & Dengue and West Nile viruses &  Antiflavivirus 4G2 antibody & 10~Plaque-Forming Units (PFU)/mL &~\cite{paul2015}\\
\\
\begin{tabular}[c]{@{}c@{}} $SiO_{2}$/Au particles (4~nm/100~nm)\\ [6pt] \end{tabular} & Zika virus &  anti- Zika (NS1) antibody & 10~ng/mL &~\cite{camacho2018}\\
\\
\begin{tabular}[c]{@{}c@{}} Ag particles (20~nm~-~80~nm)  \\ [6pt] \end{tabular}  & Dengue virus & NS1 antibody  & 0.06~$\mu$g/mL &~\cite{Pearlson} \\
\\
\begin{tabular}[c]{@{}c@{}} Au particles (40~nm)  \\ [6pt] \end{tabular}  & SARS CoV-2 & Nucleocapsid (N) protein  & 150~ng/mL &~\cite{behrouzi} \\
\\
\hline \hline
\end{tabular}}%
\end{threeparttable}
}
\end{table*}
%%%%%%%%%%%%%%%%%%%%%%%%%%%%%%%%%%%%%%%%%%%%%%%%%%%%%%%%%%%%%%%%%%%%%%%%%%%%%%

\section{Plasmonic biosensors}
Plasmonic optical biosensor technology has emerged as a powerful diagnostic tool~\cite{Sabine, Liu2020, Liyanage2022, Giovanni2022}. By selecting the appropriate biorecognition element, the technology can be applied to virtually any type of target molecule, from proteins, nucleic acids, bacteria, and drugs to human cells
~\cite{Makropoulou, Soler,Homola,Ahmet,Li},with many studies having demonstrated its utility in the biomedicine and environmental fields~\cite{QU,Pang,Zhou,Anderson,Kolomenskii}. In medicine, the accurate diagnosis of specific diseases is key for the timely and appropriate treatment and clinical management of a patient. Moreover, the rapid and early identification of certain diseases before the appearance of external symptomatology can be also important. This is the case with COVID-19; the availability of plasmonic biosensors for the rapid and accurate detection of severe acute respiratory syndrome coronavirus 2 (SARS-CoV-2) may be useful for massive population screening, the early detection of infected patients, and a more efficient management of the pandemic~\cite{Soler,Giovannini,Asghari,wang2022}. Owing to the versatility of plasmonic biosensors, the detection process can be modified. For example, the use of genomic RNA sequences of the virus target, instead of viral antigens, has enabled the rapid development of specific reverse transcription (RT)-PCR-based genomic assays~\cite{Soler}. Hence, plasmonic biosensors can be applied to the direct and label-free detection of viral RNA by designing and immobilizing single-stranded DNA probes, as receptor, with complementary sequences to specific SARS-CoV-2 gene fragments on the sensor surface~\cite{Soler}. Moreover, the sensitivity and specificity for SARS-CoV-2 could be increased with the combination of several probes targeting genes of the same virus~\cite{Soler}. Henceforth, this review we will discuss recent plasmonic biosensor platforms for virus detection, with an emphasis on SARS-CoV-2 (Table~\ref{tab:Table1}).  

\subsection{Biosensing using plasmonic nanostructures}
\begin{figure*}[h!]
\centering
\includegraphics[trim={0cm 0cm 0cm 0cm},clip, width=0.9\textwidth]{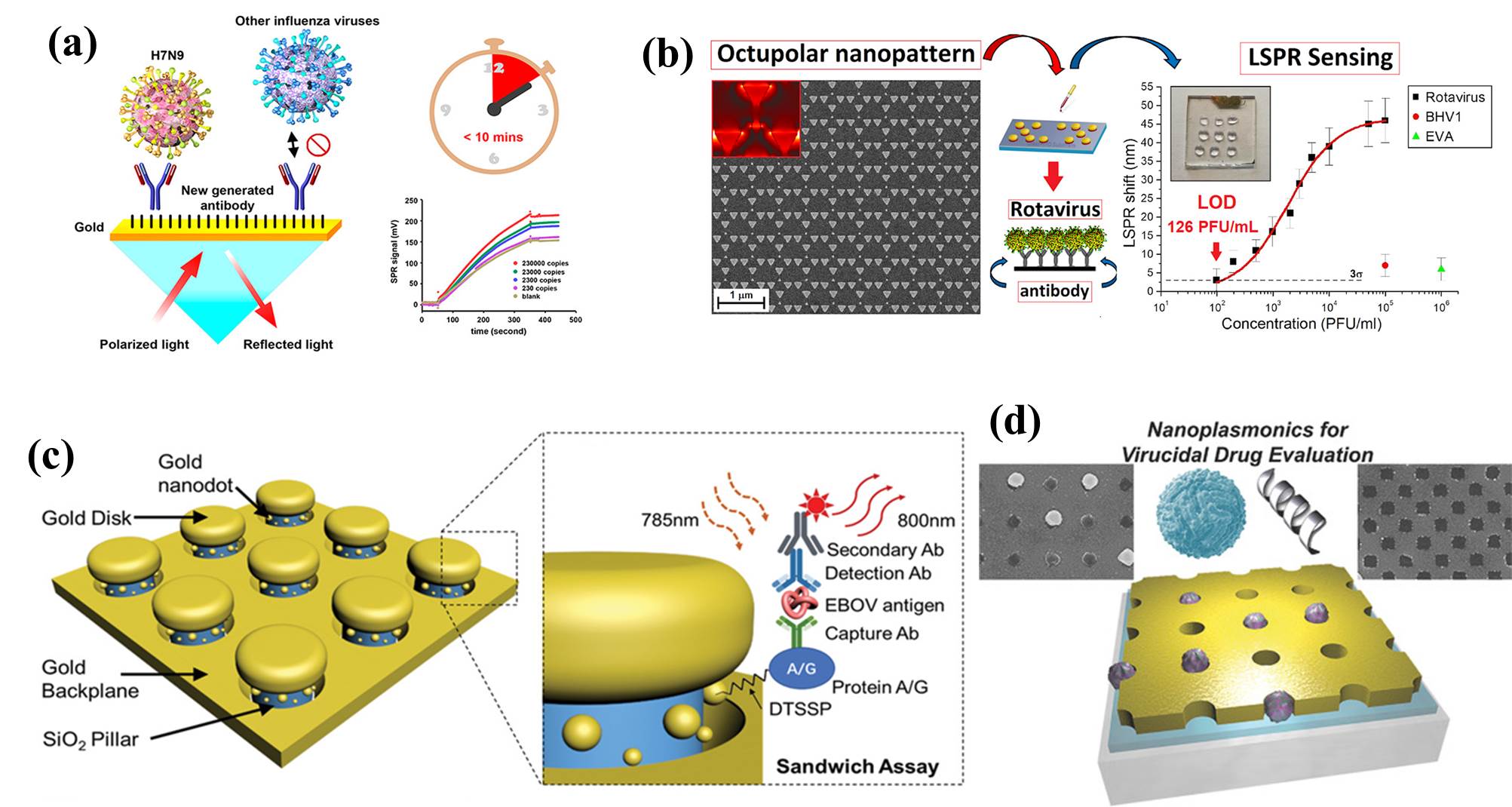}
\setlength\abovecaptionskip{10pt}
\caption{\label{Fig.4} (a) Schematic of a plasmonic biosensor used to identify avian influenza H7N9. A bare Ag/Au chip is cleaned before surface functionalization with self-assembled monolayers. The capture antibody, at a concentration of 10~$\mu$g/mL, is covalently immobilized to the reaction spot of the SPR chip (reproduced with permission from~\cite{Chang}). (b) Left: Scanning electron microscopy images of the octupolar geometry-based Au nanostructure used for rotavirus detection. The minimum interparticle distance between two unit cells is 25 nm. Right: Average LSPR peak shift (black square) and Langmuir isotherm fitting (red line) for various concentrations of rotavirus in distilled water~\cite{Rippa}).(c) Schematic of a nano-antenna array for Ebola virus detection. The gold nanodisks and backplane are separated by Si$O_{2}$ nanopillars, forming nanocavities. Gold nanoparticles are present on the Si$O_{2}$ pillar surfaces, where the localized electromagnetic field around nanostructure is highest (reproduced with permission from~\cite{Faheng}). (d) Schematic of a periodic gold nanohole array that was designed in order to selectively capture lipid vesicles and virus particles inside the nanoholes. The 10~$\times$~10~mm gold nanohole array was formed on a glass substrate by the template-stripping method. An optical adhesive layer is present between the gold and glass (reproduced with permission from~\cite{Jackman}).}  
\end{figure*}

The first implementation of plasmonic label-free biosensors for influenza virus detection was reported 25 years ago~\cite{Schofield}. Since then, researchers have developed a variety of biosensor assays for rapid virus detection and quantification based on plasmonic technologies~\cite{Ravina}. Chang~\textit{et al.} reported a simple strategy for avian influenza A (H7N9) virus detection using an intensity-modulated SPR biosensor integrated with a monoclonal antibody (Fig.~\ref{Fig.4}(a))~\cite{Chang}. Specifically, the authors employed a Kretschmann configuration using an Ag/Au (35~nm/10~nm) chip to increase the selectivity for the virus. They noted an LOD of 144 copies/mL, which indicated a sensitivity 20-fold higher than with target-captured ELISA using antibodies and better than conventional RT-PCR tests~\cite{Chang}. Furthermore, they evaluated their configuration using mimic clinical specimens containing the H7N9 virus mixed with nasal mucosa from patients with flu-like symptoms and noted a detection limit of 402 copies/mL, which was far superior to conventional influenza detection assays, and a rapid testing time of under 10 min~\cite{Chang}. Likewise, Prabowo~\textit{et al.}~\cite{Prabowo1} demonstrated a portable SPR biosensor for the quantification of enterovirus antibodies, which showed a detection limit of 67~copies/mL. In another study, an SPR-based biosensor fabricated for nine common respiratory viruses showed an LOD of 2~nM for SARS~\cite{Shi}, and an SPR chip developed to detect the Ebola virus showed a sensitivity of 0.5~pg/mL~\cite{Sharma}. The authors modified a gold chip with 4-mercaptobenzoic acid and used three monoclonal antibodies of Ebola virus to study the efficiency based on the affinity constant~\cite{Sharma}. 

%%%%%%%%%%%%%%%%%%%%%%%%%%%%%%%%%%%%%%%%%%%%%%%%%%
%Figure 4
%%%%%%%%%%%%%%%%%%%%%%%%%%%%%%%%%%%%%%%%%%%%%%%%%%
%%%%%%%%%%%%%%%%%%%%%%%%%%%%%%%%%%%%%%%%%%%%%%%%%%%%%%%

A biosensing platform developed by Diao~\textit{et al.}~\cite{Diao} based on the Biacore X analytical system was able to obtain 48~fM of HIV-1-related DNA using entropy-driven strand displacement reactions (ESDRs) as an isothermal, label-free nucleic acid amplification technique. The authors developed a sensitive SPR biosensing strategy for enzyme and label-free detection based on DNA nanotechnology~\cite{Diao}. The whole detection process was accomplished in 60~min and with high accuracy and reproducability~\cite{Diao}. The authors noted that the observed biosensing performance could be attributed to the perfect combination of a hairpin probe, ESDR circuit, and DNA tetrahedrons on the SPR biosensing chip~\cite{Diao}. Another SPR device has been developed to detect the tuberculosis (TB) virus with an LOD of 10~ng/mL using antibody responses to the Ag85-TB secretory protein~\cite{Trzaskowski}. This configuration was comparable to a commercial benchtop SensiQ Discovery SPR system and was validated in real tuberculosis patient samples~\cite{Trzaskowski}. In another study, an organic light-emitting diode (LED) prism-coupled SPR sensor was shown to have an LOD of 63~pg/mL for insertion sequence 6110, a mobile element specific for the Mycobacterium tuberculosis complex~\cite{Prabowo}. The authors quantified digoxigenin-labeled PCR products of the DNA target using SPR sensing by fabricating an SPR chips on a BK7 glass slide coated with Cr/Au (2~nm/50~nm) metal layers~\cite{Prabowo}. The authors noted that the use of LEDs as a lightweight alternative to laser-based or halogen lamp systems enabled the platform to be portable~\cite{Prabowo}.

A two-dimensional octupolar geometry-based gold nanostructure was fabricated by Rippa~\textit{et al.}~\cite{Rippa} to detect ultrasmall concentrations of rotavirus, which is the main cause of childhood viral gastroenteritis in humans (Fig.~\ref{Fig.4}(b)). Specifically, the authors designed an array of units comprising three large identical triangular gold nanoprisms (side length, 200~nm) and one smaller inner prism (side length, 80~nm), with a 25~nm separation between adjacent units~\cite{Rippa}. An LOD of 126~$\pm$~3~PFU/mL plaque forming units (PFU)/mL using a very low sample volume (2~$\mu$L) was estimated. In addition, the authors evaluated their plasmonic biosensor with two more viruses (bovine herpesvirus [BHV1] and equine viral arteritis [EVA]) to confirm its sensitivity and specificity. A maximum LSPR peak shift of 7~nm from a concentration of 1~$\times$ $10^{5}$ PFU/mL for BHV1 was measured, while a lower maximum shift from a lower concentration was observed (6 nm) for EVA~\cite{Rippa}.

Recently, an array of gold nano-antenna that uses a sandwich immunoassay format has been fabricated for single-molecule detection of Ebola virus antigens (Fig.~\ref{Fig.4}(c))~\cite{Faheng}. The nano-antenna consists of SiO$_{2}$ nanopillars bound to gold nanodisks and nanodots, which enhance the fluorescence signal through the formation of nanocavities~\cite{Faheng}. The authors used a thiol-gold link and a protein A/C layer  to simultaneously functionalize the surface of the nanopillars and to prevent signal losses on the gold surfaces~\cite{Faheng}. They noted a detection sensitivity for the Ebola virus soluble glycoprotein in human plasma of 220~fg/mL; this was a significant improvement over the recommended immunoassay test for Ebola virus antigens~\cite{Faheng}. In addition, the interaction of light with the periodic array of nanoholes enabled the extraordinary optical transmission effect~\cite{bethe}, which enhanced the transmission of light at specific wavelengths. These spectral characteristics have facilitated the development of high-sensitivity plasmonic biosensors that can be integrated with microfluidics.
A metallic nanohole-based assay was developed~\cite{Jackman} to capture single virus-like particles (Fig.~\ref{Fig.4}(d). The diameter of the nanoholes was chosen to fit the size distribution of virus particles that had been treated with a virucidal drug candidate~\cite{Jackman}. The sensing performance of the platform was evaluated by monitoring resonance shifts for the virucidal-induced capture of single virus-like particles, showing a minimum RI resolution of
5.5~$\times$ $10^{-5}$ RIU~\cite{Jackman}. The authors noted high RI sensitivity in the functionalized nanoholes with a low surface coverage when compared with non-functionalized nanoholes~\cite{Jackman}.

%%%%%%%%%%%%%%%%%%%%%%%%%%%%%%%%%%%%%%%%%%%%%%%%%%
%Figure 5
%%%%%%%%%%%%%%%%%%%%%%%%%%%%%%%%%%%%%%%%%%%%%%%%%%
\begin{figure*}[ht]
\centering
\includegraphics[trim={0cm 0cm 0cm 0cm},clip, width=1\textwidth]{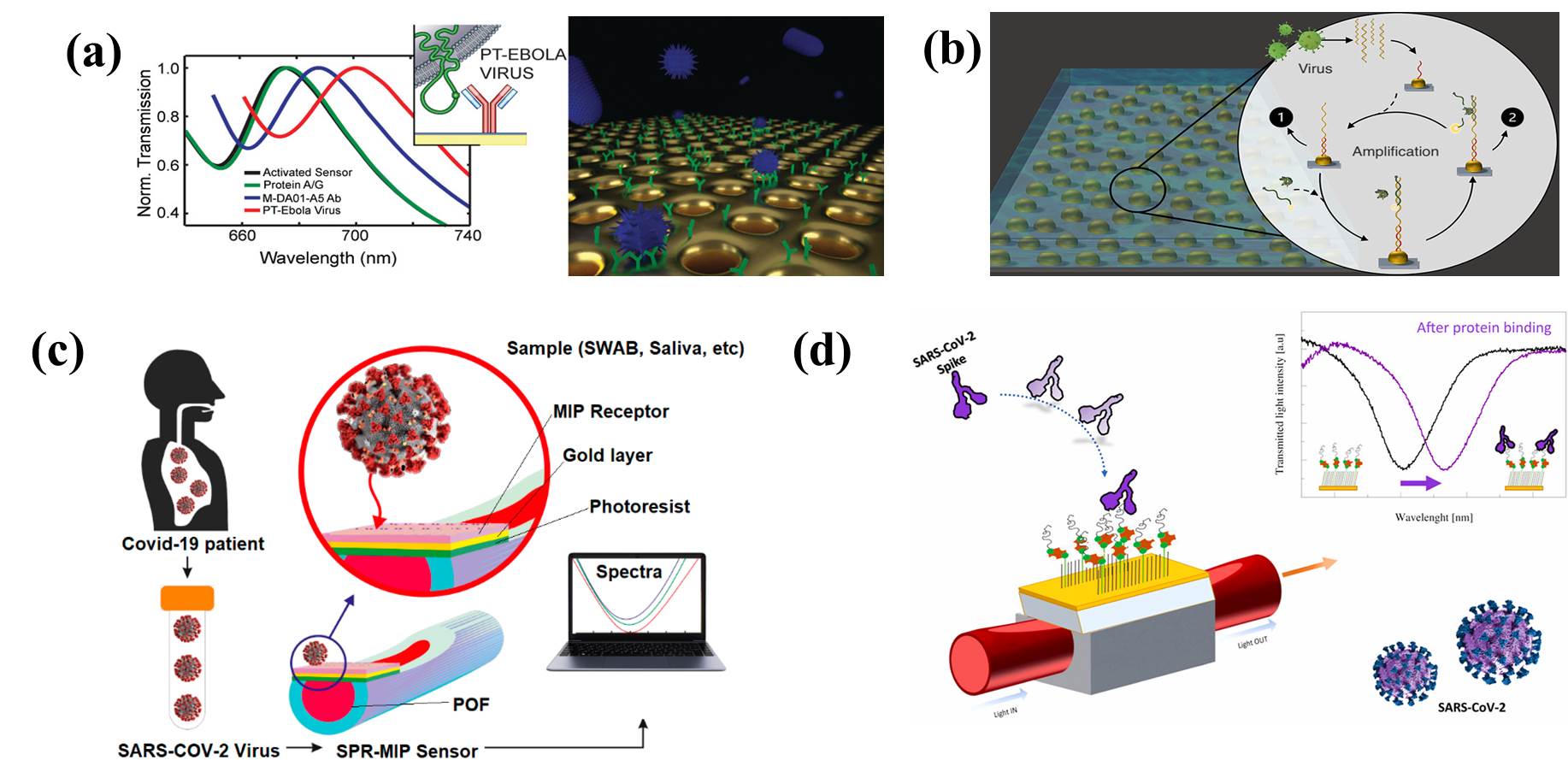}
\setlength\abovecaptionskip{10pt}
\caption{\label{Fig.5}(a) SSchematic of the thermoplasmonic-assisted dual-mode approach. Amplification-free-based direct viral RNA detection and amplification-based cyclic fluorescence probe cleavage detection are combined to provide SARS-CoV-2 detection within 30~min (reproduced with permission from~\cite{Yanik}). (b) A label-free optofluidic nanoplasmonic sensor that can detect vesicular stomatitis virus and pseudotyped Ebola virus from biological media with little to no sample preparation (reproduced with permission from~\cite{Qiu1}).(c) Schematic of a biosensor based on a plasmonic plastic optical fiber coupled with a novel type of synthetic molecularly imprinted polymer (MIP) receptor for the specific recognition of subunit 1 of the SARS-CoV-2 spike (S) protein (reproduced with permission from~\cite{Cennamo}). (d) Biosensing configuration based on an SPR D-shaped plastic optical fiber integrated with an aptamer for the recognition of the receptor-binding domain (RBD) of the S glycoprotein of SARS-CoV-2~\cite{Cennamo2}).}  
\end{figure*}
%%%%%%%%%%%%%%%%%%%%%%%%%%%%%%%%%%%%%%%%%%%%%%%%%%%%%%%

Since terahertz waves are non-ionizing and harmless to organic tissues and biomolecules, they may become increasingly attractive for biomedical applications~\cite{Yu1}. A terahertz gold metasensor was designed for Zika virus envelope protein detection~\cite{Ahmadivand1}. Based on toroidal metamaterial properties, these devices support resonances that possess much higher sensitivity to RI perturbations in the surrounding media~\cite{Pendry,Smith}. The toroidal metamaterial consisted of an array of mirroring asymmetric split resonators and had the ability to support a \textit{Q}-factor~\cite{Ahmadivand1} around 30. By measuring the shift of the toroidal dipolar momentum, the authors determined the LOD and sensitivity of the metasensor to be 560~pg/mL and 5.81~GHz/log(pg/mL), respectively, for a variety of Zika virus concentrations~\cite{Ahmadivand1}.  

\subsubsection{Plasmonic nanostructures for SARS-CoV-2 detection}

Ahmadivand~\textit{et al.} demonstrated femtomolar-level detection of the SARS-CoV-2 S protein using toroidal gold metasensors~\cite{Arash1}. The authors improved the binding properties of the device by functionalizing gold nanoparticles with antibodies for the S protein, resulting in an LOD of 4.2 fmol/mL~\cite{Arash1}. A low-cost plasmonic sensor consisting of an Au-TiO$_{2}$-Au nanocup array was demonstrated that permitted observation of the plasmon resonance wavelength and intensity change of S protein capturing events by utilizing the extraordinary optical transmission effect in transmission light spectroscopy~\cite{Huang}. The authors achieved an LOD of 370~virus particles (vp)/mL with a virus concentration in the range of 0–10$^{7}$~vp/mL

A plasmonic nanohole array with S protein antibodies immobilized on the surface was fabricated~\cite{Yanik} to detect a broad range of pathogens in a typical biology laboratory setting (Fig.~\ref{Fig.5}(a)). By capturing the S proteins, whole virus particles could be suspended in the nanohole array, which resulted in a red-shift of the resonance~\cite{Yanik}. A plasmonic microfluidic biosensing platform was developed by Funari~\textit{et al.} who demonstrated the utility of electrodeposition-based gold nanospikes combined with optical probes~\cite{Funari}. Based on local RI changes caused by the interaction of the SARS CoV-2 S protein and antibodies in the diluted human serum, a shift of the LPRS resonance peak was detected, with a detection concentration of 0.08~ng/mL~\cite{Funari}. The authors noted that the proposed platform could complement existing serological assays and improve COVID-19 diagnosis~\cite{Funari}. A dual functional plasmonic detection platform that combines the plasmonic photothermal~\cite{sile2} and LSPR effects has been reported for SARS-CoV-2 detection~\cite{Qiu,Qiu1} (Fig.~\ref{Fig.5}(b)). Two-dimensional gold nano-islands functionalized with RdRp-COVID cDNA (RdRp-COVID-C) receptors permit the selective detection of RdRp-COVID-C through DNA hybridization, giving an LOD for the cDNA of 0.22~pM. This provides a new approach for SARS-CoV-2 detection~\cite{Qiu,Qiu1}. 

\subsection{Biosensing using plasmon-based optical fiber}

In the past few decades, optical fibers have evolved from an optical transmission waveguide to important components of applications ranging from small particle manipulation~\cite{fiber4,fiber2,fiber1,fiber3,addanki2018} to medical imaging~\cite{Podoleanu}.
The past decades, a new class of optical fiber sensors based on SPR has been added to the family of PoC devices~\cite{Esfahani}. Plasmonic fiber-optic biosensors offer an interesting alternative to classical prism-based configurations and are advantageous in terms of flexibility and cost. Plasmonic optical fiber platforms have provided miniaturized sensing approaches for the determination of clinical biomarkers~\cite{Gauglitz}.

Of particular note is an SPR-based optical fiber device that has been developed for the analysis of avian influenza virus subtype H6~\cite{Xihong}. The SPR-based optical fiber consists of a 40~nm thin gold film and a side-polished structure~\cite{Xihong}. To optimize the self-assembled monolayers and subsequent antibody functionalization, the detection surface of the SPR-based optical fiber was modified with plasma at low temperature, which rendered better results than chemical modification~\cite{Xihong}. The binding interaction between immobilized antibodies and antigens on the cell surface was evaluated with $10^{4}$ to $10^{8}$ embryo infectious dose (EID) 50/0.1~mL of virus, leading to a detection limit of 5.14$\times$$10^{5}$ EID$_{50}$/0.1~mL and an average response time of 10~min~\cite{Xihong}. The combination of the optical properties of LSPR nanostructures with the total internal reflection of optical fiber configurations can provide signal enhancement and better spatial sensitivities. The integration of gold nanorods into a fiber-optic platform permitted the development of an immunosensor for the determination of Cymbidium mosaic virus and Odontoglossum rings spot virus~\cite{Hsing} in plants. To achieve direct sensing of the analytes, gold nanorods were employed to generate a near-infrared sensing window to solve the color interference issue of sample matrices~\cite{Hsing}. The LOD of viral antigens after the gold nanorods of the fiber optic LSPR platform~\cite{Hsing} had been functionalized with antibodies was 48~pg/mL, while the RI resolution was 8$\times$10$^{-6}$ RIU. The authors noted that the improvement in sensitivity in comparison with ELISA was attributed to the properties of nanorods, which simultaneously prevented the color interference of similar-sized nanospheres~\cite{Hsing}. A tilted fiber grating surface coated with gold nanoparticles has been demonstrated for the detection of Newcastle disease virus (NDV)~\cite{Luo}. Modification of the fiber cladding with gold nanoparticles (with an average diameter of 80~nm) enhanced sensitivity as a result of the LSPR field, while activation of the nanoparticles with staphylococcal protein A improved the bioactivity of anti-NDV monoclonal antibodies by up to ten times compared with that of a tilted fiber grating without gold nanoparticles~\cite{Luo}. Monitoring of resonance wavelength red-shifts showed a minimum detectable amount for virus of approximately 5~pg; this is slightly better than that achievable by RT-PCR (10~pg).  

\subsubsection{Plasmonic optical fibers for SARS-CoV-2 detection}

During the recent pandemic, a plasmonic fiber optic absorbance biosensor was successfully fabricated to detect the SARS-CoV-2 nucleocapsid (N) protein~\cite{Marugan}. The integration of gold nanoparticles into a multimode U-bent optical fiber permitted the detection the N protein in a patient's saliva sample within 15~min~\cite{Marugan}. An alternative biorecognition system based on aptamers immobilized on gold nanorods embedded on D-shaped optical fibers was also fabricated for the detection the SARS-CoV-2 S protein (Fig.~\ref{Fig.5}(c))~\cite{Cennamo}. The viral protein was detectable at an LOD of 37~nM  and resonance shifts of 3.1~nm, thereby providing the ability to detect small viral concentrations~\cite{Cennamo}. The same group fabricated a synthetic MIP receptor, which was incorporated into a 60~nm thick gold film D-shaped optical fiber, for the identification of SARS-CoV-2 (Fig.~\ref{Fig.5}(d))~\cite{Cennamo2}. In this work~\cite{Cennamo2}, the authors noted that the sensitivity of the proposed plasmonic biosensor was higher than RT-PCR and with a response time of about 10 min. A photonic quasi-crystal fiber (PQCF)-based plasmonic platform was designed  to provide a theoretical sensitivity of 1,172~nm/RIU for the detection of SARS-CoV-2 within saliva~\cite{Aliee}. The PQCF consisted of 280~nm diameter air holes and a 300~nm diameter gold ring around one air hole near the core of a lattice with a 500~nm period~\cite{Aliee}. Biosensing and the amplification of targeted analytes at low concentrations are among the most important properties of a sensor for the detection of analytes. Therefore, Saad~\textit{et al.} analyzed the sensitivity and resolution of an optical fiber-based system with gold-silver alloy nanoparticles embedded in its core and covered by a layer of graphene~\cite{Saad}. The authors~\cite{Saad} showed that the system had a maximum sensitivity of 7,100 nm/RIU, FOM of 38.8 RIU$^{-1}$, and signal-to-noise ratio of 0.38. Moreover, Wu~\textit{et al.}~\cite{Wu} showed that the combination of metallic nanostructures with graphene can provide better biological sensing because of the adsorption of analytes to the graphene through  $\pi$~-~$\pi$ stacking. Hence, the modification of the plasmonic optical fiber with graphene layers may improve further the performance and detection ability of the future biosensors. 

%%%%%%%%%%%%%%%%%%%%%%%%%%%%%%%%%%%%%%%%%%

\subsection{Sensing using plasmonic nanoparticles}

The characteristics of noble metal nanoparticles have found greatest use in LSPR biosensing, with several applications utilizing this technique~\cite{Wei,Ibrahim,Geng,wang2022}. For example, a sandwich immunoassay with gold nanoparticles LSPR chip format was developed  to detect the hepatitis B virus (HBV) surface antigen (HBsAg)~\cite{Jinwoon}. For this purpose, a glass substrate was fabricated with synthesized gold nanoparticles (AuNPs) of three different sizes (15, 30, and 50 nm) and conjugated with an anti-HBsAg antibody to detect the target antigen~\cite{Jinwoon}. After 10 min, a second layer of AuNPs conjugated with the anti-HBsAg antibody was added to obtain a hetero-assembled chip, the LOD of which was 100~fg/mL~\cite{Jinwoon}. 
Takemura~\textit{et al.} used the LSPR signal from Ab-conjugated thiol-capped AuNPs to amplify the fluorescence intensity signal of quantum dots for the detection of nonstructural protein 1 (NS1) of the Zika virus~\cite{Takemura}. Their biosensor had a wide detection range of 10-10$^7$ RNA copies/mL and maintained its specificity with human serum~\cite{Takemura}. Chowdhury~\textit{et al.} also developed a biosensor using AuNPs and CdSeTeS quantum dots to identify the serotypes of dengue virus~\cite{chowdhury}. The biosensor had an LOD at the femtomolar level and was successfully applied to RNA extracted from dengue virus culture fluids~\cite{chowdhury}. Lee~\textit{et al.} constructed a label-free biosensor for avian influenza virus (H5N1) using hollow spike-like AuNPs and a multifunctional three-way DNA junction~\cite{lee2019}. To achieve the multifunctionality, each piece of DNA was tailored to aptamers specific for the hemagglutinin (HA) protein of the virus, fluorescence dye, and thiol group. The sensor detected the HA protein of H5N1 in phosphate buffered saline and chick serum with an LOD of 1~pM~\cite{lee2019}. Heo~\textit{et al.} fabricated gold nanoparticles with an approximate average size of 20~nm to detect human norovirus~\cite{Heo}. The authors’ novel sensing approach utilized noroviral protein-recognizing affinity peptides, which are relatively cost-effective compared with antibodies, to bind noroviral proteins~\cite{Heo}. They noted an LOD of the capsid protein of 9.9~copies/mL~\cite{Heo}.

AuNPs or roughened gold surfaces are also widely used in SERS spectroscopy because of their LSPR properties~\cite{wang2022}. Paul~\textit{et al.} developed an antibody-conjugated AuNP-based SERS probe for the identification of mosquito-borne viruses~\cite{paul2015}. They successfully detected dengue virus type-2 and West Nile virus at a low concentration of 10 PFU/mL~\cite{paul2015}. Camacho~\textit{et al.} designed SERS nanoprobes using gold shell-isolated nanoparticles, which contained 100~nm gold nanoparticles and 4~nm silica shells. The silica shells modified with Nile blue functioned as the Raman reporter~\cite{camacho2018}. This configuration was irradiated with a 633~nm wavelength laser beam, and the Raman signal was recorded by a mapping process. The nanoprobes successfully detected Zika virus at a very low concentration (around 10~ng/mL) and without any cross-reactivity with dengue virus~\cite{camacho2018}. Luan~\textit{et al.} developed a stable and bright fluorescent plasmonic nanoscale construct that consisted of a bovine serum albumin (BSA) scaffold with approximately 210 IRDye 800CW fluorophores, a polysiloxane-coated gold nanorod acting as a plasmonic antenna, and biotin as a high-affinity biorecognition element~\cite{Luan}. This configuration was able to improve the LOD of fluorescence-linked immunosorbent assays by up to 4,750-fold, shorten overall assay times (to 20 min), and lower sample volumes. The authors attributed this improvement in sensitivity to the BSA blocking method, in which BSA acts as a blocking agent to minimize non-specific binding of the plasmonic fluorophore to arbitrary surfaces and biomolecules.

Compared with gold nanoparticles, silver nanoparticles display a higher efficiency of LSPR excitation and a wider wavelength range~\cite{liang2015}. Moreover, silver nanoparticles have sharper LSPR bands, are less dissipative, and perform better in SERS~\cite{Ibrahim}. However, there are fewer studies on AgNPs than AuNPs as plasmonic biosensors. One reason behind this may be that AgNPs display toxicity~\cite{kim2012, zhang581} and antiviral effects~\cite{jeremiah,salleh}. Another reason is that bare AgNPs are not as stable as AuNPs because of quicker oxidation~\cite{Wang_2018}. 
To deal with these drawbacks, AgNPs are usually coated with different materials~\cite{tejamaya}, and the coating material and thickness greatly influence the optical properties of the AgNPs. A thermally annealed thin silver film deposited onto a silicon substrate was used to detect the NS1 antigen of dengue virus in whole blood~\cite{Pearlson}. After the annealing process, silver nanoparticles with diameters from 20 to 80 nm were generated and with inter-structural spacing ranging from a few tens to about 100 nm~\cite{Pearlson}. The authors determined the system to have an RI sensitivity of $10^{-3}$,  while an increase in absorption and a red-shift of 108 nm of the peak absorption wavelength were observed with antigen binding~\cite{Pearlson}. The sensitivity of this configuration was found to be 9~nm/($\mu$g/mL), and the LOD was 0.06~$\mu$g/mL. Hong~\textit{et al.} developed hybrid slot antenna structures with silver nanowires in the terahertz frequency range to detect bacteriophage PRD1 and obtained an enhancement factor of 2.5 for a slot antenna width of 3~$\mu$m~\cite{Hong2018}. 

A SERS platform has been used to detect HBV~\cite{Lu}. The authors  used a standard, label-free Ag nanoparticle solution as the SERS-active substrate to test blood serum samples from HBV patients and healthy volunteers~\cite{Lu}. The SERS spectra of the serum samples from both the infected patients and healthy volunteers were compared by employing linear discriminant analysis~\cite{Lu}. Using this approach, a SERS spectrum was produced in 10 min for each sample and with a diagnostic sensitivity of 91.4\%, indicating the great potential of this for a quick, non-invasive, label-free diagnostic method through the implementation of principal components analysis~\cite{Lu}.

\subsubsection{Metallic nanoparticles for SARS-CoV-2 detection}

Behrouzi and Lin applied LSPR of antigen-coated AuNPs to detect the N protein of SARS-CoV-2~\cite{behrouzi}. This detection method gave naked-eye results in 5 min and with an LOD of 150 ng/mL~\cite{behrouzi}. Park~\textit{et al.} used self-assembly AuNPs arrays for the detection of the SARS-CoV-2 S protein~\cite{park2022}. Their biosensor gave quick results with high sensitivity in just 10 min and without any purification steps~\cite{park2022}. Both aforementioned sensors could be used for the PoC detection of SARS-CoV-2. In addition Das~\textit{et al.} achieved an LOD of the S protein of 111.11 deg/RIU using a gold nanorod with a Kretschmann prism configuration~\cite{Das1}. Zavyalov~\textit{et al.} built a SERS aptasensor based on hydroxylamine-reduced AgNP substrates and successfully detected SARS-CoV-2 in about 7 min and with an LOD of 5.5$\times$10$^4$ median tissue culture infectious dose/mL~\cite{zavyalov}. Tripathi~\textit{et al.} deposited AgNPs over glass coverslips and used them as SERS substrates~\cite{tripathi}. The sensor was used to detect the Japanese encephalitis virus and demonstrated ultrasensitive detection, with a detection limit of about 7.6 ng/mL and a linear response from 5 to 80 ng/mL~\cite{tripathi}. A colorimetric assay based on AgNPs with diameters less than 60 nm and capped with thiol-modified antisense oligonucleotides specific for the N protein of SARS-CoV-2 has been demonstrated as being capable of diagnosing positive COVID-19 cases from isolated RNA samples within approximately 10 min~\cite{Moitra}.

%%%%%%%%%%%%%%%%%%%%%%%%%%%%%%%%%%%%%%%%%%
\section{Perspectives and Outlook}
 
Progress in material science and fundamental optics will continue to provide advantages to biosensing research. An important facilitator of this progress is the use of numerical analysis tools, such as Comsol Multiphysics software~\cite{EPA}, to explore geometrical and material parameters for the optimization of a biosensor's performance.
The combination of various optical and non-optical, for example electrochemical, detection configurations on a single platform could also enable multifunctional biosensors to extract information from a given sample. Likewise, two-dimensional materials, such as graphene, can provide dynamic control of plasmonic resonances, which is needed for small molecule detection~\cite{Lee2015}. Another alternative technique to be considered is nanopore technology, which allows for the precise detection of subunits as well as the sequencing of pathogen DNA and RNA in an effective and versatile way; this technology will be at the forefront of future state-of-the-art approaches~\cite{Wang}. Nanopore-based sequencing systems, such as the one developed by Oxford Nanopore Technologies were successfully applied to SARS-CoV-2 strains at the early stages of the pandemic~\cite{Wei2020,Bull2020}. The rapid and real-time detection of mutagenized virus are key benefits of this technique, providing important data for further epidemiological analysis.  

In parallel, the availability of a variety of metal nanoparticle synthesis protocols as well as an increase in the number of commercial nanoparticle providers may contribute to the development of novel biosensors with high specificity and selectivity~\cite{Soler,Hatice,wang2022,Li}. To develop high-sensitivity tests for SARS-CoV-2, the selection of metallic nanoparticles with appropriate sizes and shapes is a key point since their physical and optical properties can greatly influence the performance of a nanoparticle-based diagnostic system. Although spherical nanoparticles have been studied most extensively because of their ease of synthesis, other shapes are worth investigating when a higher sensitivity or a different sensing strategy is desired but not achievable with nanospheres. Regarding nanoparticle size, large metallic nanoparticles have large absorption cross-sections and may result in systems with higher sensitivities than those utilizing small nanoparticles. However, all of these parameters need to be addressed and evaluated on a project-by-project basis as many other factors could influence the LOD~\cite{Olivier}. 

Despite the excellent biosensing performance of plasmonic diagnostic tools, several technological aspects still require considerable improvement before fully operative devices for clinical diagnosis and real-world applications can be realized. Factors that need to be addressed include cost, sensitivity, specificity, and reproducibility, as well as user interfaces and connectivity that allow for real-time monitoring of data collection. In terms of cost, inexpensive disposal chips are necessary to avoid cross-contamination issues and complicated cleaning procedures while handling biological samples or body fluids. In this regard, an integrated approach that allows for single-use cartridges and a stand-alone reader are desirable. Microfluidic technology can also play a key role in providing disposable, stable over time, and easy to manipulate cartridges through the incorporation of biochips with specific biofunctionalities for each detection assay. For airborne respiratory viruses, it will be essential to integrate such cartridges with additional safety steps, include sample preprocessing, before final detection~\cite{Cui2015}.
Another possible advancement is the merging of plasmonic devices with smartphones; their light sources, cameras, and image processing and communication capabilities can reduce costs and facilitate large scale distribution~\cite{Smart,Lopez,Qiang_smartphone, SALIMIYANRIZI2022100925}. Therefore, the development of portable and wireless nanobiosensors is essential for diverse applications. Sample collection and processing is an additional consideration for on-site biosensing. The large diversity of analytes and the matrix composition of specimens such as body fluids still remain a challenge. For example, virus detection from clinical specimens is still limited owing to the lack of proper methods to prevent the interference of biomolecules in body fluids. In this sense, the design of antifouling coatings that can take into account either the composition of the media or the biological receptor's characteristics may help to bridge the gap between common analytical methods and plasmonic biosensing applications. 

In summary, this tutorial review highlighted the physics underpinning the mechanics of plasmonic-based biosensors, the current progress of biosensor research, and the ability of such devices to detect viruses. It is worth noting that although high sensitivity is always the main goal of any biosensor, for better clinical and commercial translation, it is essential to balance the trade-off between the sensitivity, cost-effectiveness, portability, and stability of these plasmonic-based systems. Against the backdrop of the COVID-19 global pandemic, continued biosensor development is crucial for the realization of more portable and affordable platforms that can meet global healthcare needs.  

\section*{Acknowledgements}
This work was supported by funding from  Duke Kunshan University and Synear and Wang-Cai Grant(DKU). 
%DGK acknowledges support from JSPS Grant-in-Aid for Scientific Research (C) Grant Number GD1675001, the OIST and DKU Editing Section for reviewing the manuscript.

\section*{Conflicts of interest}
The authors declare that they don't have any competing financial interests or personal relationships that could have appeared to influence the work reported in this tutorial review.
%%%END OF MAIN TEXT%%%

%  For footnotes in the main text of the article please number the footnotes to avoid duplicate symbols. e.g.  \footnote[num]{your text} the corresponding author \ast counts as footnote 1, ESI as footnote 2, e.g. if there is no ESI, please start at [num]=[2], if ESI is cited in the title please start at [num]=[3] etc. Please also cite the ESI within the main body of the text using \dag.

% The \balance command can be used to balance the columns on the final page if desired. It should be placed anywhere within the first column of the last page.

% \balance

% If notes are included in your references you can change the title from 'References' to 'Notes and references' using the following command:
% \renewcommand\refname{Notes and references}

%%%REFERENCES%%%
\scriptsize{
\bibliography{Bibliography} %You need to replace "rsc" on this line with the name of your .bib file
\bibliographystyle{rsc} } %the RSC's .bst file

\end{document}